\documentclass[]{iopart}
\pdfoutput=1
\usepackage{graphicx}

\begin{document}

\paper[Verhulst-like two-species population dynamics model]{Full analytical solution and complete phase diagram analysis of the Verhulst-like two-species population dynamics model}

\author{Brenno Caetano Troca Cabella$^1$, Fabiano Ribeiro$^2$ and Alexandre Souto Martinez$^{1,3}$}

\address{$^1$ Faculdade de Filosofia, Ci\^encias e Letras de Ribeir\~ao Preto (FFCLRP) \\
               Universidade de S\~ao Paulo (USP) \\
               Av.~Bandeirantes, 3900 \\
               14040-901  Ribeir\~ao Preto, SP, Brazil}

\address{$^2$ Departamento de Ci\^encias Exatas (DEX), \\
                Universidade Federal de Lavras (UFLA) \\
	           Caixa Postal 3037 \\
		     37200-000 Lavras, MG, Brazil}

\address{$^3$ National Institute of Science and Technology in Complex Systems
               }
\eads{\mailto{brenno@usp.br}, \mailto{flribeiro@dex.ufla.br}, \mailto{asmartinez@usp.br}} 

\begin{abstract}
The two-species population dynamics model is the simplest paradigm of interspecies interaction.   
Here, we include intraspecific competition to the Lotka-Volterra model and solve it analytically. 
Despite being simple and thoroughly studied, this model presents a very rich behavior and  some characteristics not so well explored, which are unveiled. 
The forbidden region in the mutualism regime and the dependence on initial conditions in the competition regime are some examples of these characteristics.  
From the stability of the steady state solutions, three phases are obtained: (i) extinction of one species (Gause transition), (ii) their coexistence and (iii) a forbidden region. 
Full analytical solutions have been obtained for the considered ecological regimes. 
The time transient allows one to defined time scales for the system evolution, which can be relevant for the study of tumor growth by theoretical or computer simulation models. 
\end{abstract}

\submitto{Journal of Statistical Mechanics}

\section{Introduction}

The ecological community is composed by a complex interaction network from which the removal of a single species may cause dramatic changes throughout the system.  
The interactions between species only became better formalized in population dynamics with the Lotka-Volterra equations, in the 1920's~\cite{murray}.
This simple theoretical model explains prey-predator behavior and others range  of natural phenomena, for instance, the oscillatory behavior in a chemical concentration in chemistry or in fish catches in ecology, cells of the immune system and the viral load (in immunology)~\cite{nowak} econophysics \cite{solomon_2000} etc.

Besides the predation interaction described by Lotka-Volterra equations, there are many other different kinds of interaction taking place between biological species. 
If species unfavor each other, for instance,  when two species occupy the same ecological niche and use the same resources, there is \textit{competition}.  
If species favor each other, as in pollination/seed dispersion by insects, there is \textit{mutualism} - also called  \textit{symbiosis}.  
If only one of these species is independent of the other, there are two possibilities: \textit{amensalism}, if the considered species has a \textit{negative} effect on the other; and \textit{commensalism}, otherwise. 
As an example of amensalism, consider that an organism exudes a chemical compound as part of its normal metabolism to survive, but this compound is detrimental to another organism. 
An example of commensalism is the remoras that eat leftover food from the shark. 
Finally, when species do not interact at all there is the so-called \textit{neutralism}. 

The Verhulst-like two-species population model incorporates limit environmental resources, logistic growth of one species in the absence of the other, and interspecific interaction. 
It is described by the following equations~\cite{murray,Keshet}: $dN_1/dt = \kappa_1 N_1 ( 1 - N_1/K_1 + \alpha_1 N_2/K_1 )$ and $dN_2/dt = \kappa_2 N_2 ( 1 - N_2/K_2 + \alpha_2 N_1/K_2 )$, where $N_i \ge 0$, $\kappa_i$ and $K_i > 0$ are the number of individuals (size), net reproductive rate, and the carrying capacity of species $i$ ($=1,2$),  respectively. 
The term $-\kappa_1 N_1^2/K_1$ represents the competition between individuals of the same species (intraspecific competition), and $-\kappa_1 N_1 N_2/K_1$ represents the interaction between individuals of different species (interspecific interaction).
The carrying capacity  represents the fact that both species are resource-limited. 
In fact, $K_1$ represents the restriction on resources that comes from any kind of external factors, but the ones relative to species 2, similarly for $K_2$. 
To use non-dimensional quantities in the Verhulst-like two-species model, write $p_i = N_i/K_i \ge 0$, for $i=1,2$. 
Time is measured with respect to the net reproductive rate of species 1, $\tau = \kappa_1 t \ge 0$. Here,  we restrict ourselves to the case  $\kappa_i>0$.
The scaled time is positive since we take the initial condition as $t_0 = 0$. 
Moreover, the two net reproductive rates form a single parameter  $\rho = \kappa_2/\kappa_1 > 0$, fixing a second time scale to the system:  $\tau' \equiv \rho \tau = \kappa_2 t$. 
The parameters $\alpha_1$ and $\alpha_2$ represent the interaction degree and also the kind of interaction between species.
The non-dimensional population interaction parameters are given by  $\epsilon_1 = \alpha_1 K_2/K_1$ and  $\epsilon_2 = \alpha_2 K_1/K_2$, which are not restricted and represent the different ecological interactions. 
With these quantities, the Verhulst-like two-species model become:   
\begin{eqnarray}
 \frac{d p_1}{d\tau} &=&          p_1[ 1 -p_1 + \epsilon_1 p_2] = f(p_1,p_2)  \label{simple-model_1} \\
 \frac{d p_2}{d\tau} &=&  \rho  p_2[ 1- p_2 + \epsilon_2 p_1] = g(p_1,p_2) \label{simple-model_2} \; .
\end{eqnarray}
Contrary to $\rho$, which has no major relevance to this model (since we consider only $\rho>0$), the product $\epsilon_1\epsilon_2$ plays an important role, so that $\epsilon_1 \epsilon_2 < 0$ means predation;  $\epsilon_1\epsilon_2 = 0$ means commensalism, amensalism, or neutralism; and  $\epsilon_1 \epsilon_2 > 0$ means either mutualism or competition as depicted in the diagram of  \Fref{phase-diagram}.

The novelty of the present work is two-fold. 
We give a new interpretation of the Verhulst-like two-species model, since we do not restrict the interaction parameter $\epsilon_i$, as it is usually done in previous studies~\cite{Goel:1971p1001,Dorschner:1987p2927,sigmund}.
This  absence of restriction on the interaction parameters allows us to display many different ecological regimes (competition, predation, and mutualism) in the same phase diagram, which are obtained by the steady state solutions~\cite{goh_1976,hastings_1978,steinmuller_1982,myerscough_1992,gleria_2005,hofbauer_2008}.  
Nevertheless, we point out the existence of a survival/extinction transition, as well as a transition for a forbidden region in the mutualism regime.  
Also, based on the result of Ref.~\cite{brenno_2011}, we have been able to find the complete analytical solution for this model. 
From the analytical solutions, the transient time allows one to establish characteristic times~\cite{hastings_2004,gavrilets_1995,lai_1995,lai_1995_2,lai_1995_3,cavalieri_1995,harrison_2001,kaitala_1999}
 which can be relevant, for instance for tumor growth models~\cite{wodarz_2001, wodarz_2005,novozhilov_2006}.
Also transient solutions may be used as a guide to validate numerical simulation algorithms, giving to Monte Carlo time steps a real time scale ~\cite{shabunin_2008}.
Furthermore the analytical solutions may be used in stochastic theoretical modeling of two-species systems to give insight when averages are done ~\cite{reichenbach_2006}.


The text is organized as follows. 
In Sec.~\ref{sec:twospecies_pdm}, the steady state solution of the interspecific competition in the Lotka-Volterra equations is presented. 
These solutions correspond to the stable ecological regimes in the parameter space diagram, where we point out the existence of the survival/extinction transition and of a non-physical region in the mutualism regime.  
We show that although our model is simpler, it is able to reproduce the four scenarios found in more complete tumor growth models.
In Sec.~\ref{neutralism-amensalism-comensalism}, the analytical solutions for the trivial case neutralism and the non-trivial ones amensalism and comensalism are presented.
Next, the full analytical solutions are obtained for the mutualism, predation, and competition regimes. 
Our conclusions are described in Sec.~\ref{conclusion}.

\section{Steady state solutions}
\label{sec:twospecies_pdm}

To obtain the steady state solution $p^*_1 = p_1(\tau \to \infty)$ and $p_2^*=p_2(\tau \to \infty)$ of \Eref{simple-model_1} and \Eref{simple-model_2}, we have to impose $d p_1/d\tau = d p_2/d \tau = 0$, which implies  $f(p^*_1,p^*_2) = g(p^*_1,p^*_2) = 0$ and  leads to $p_1^*(1-p_1^* + \epsilon_1 p_2^*)  =  0$ and  $p_2^*(1-p_2^* + \epsilon_2 p_1^*)  =  0$. 
One has the following trivial (``t''), semi-trivial (``st''), and non-trivial (``nt'') pairs of solutions: (i) $p_{1,  t}^*    = 0 $ and $ p_{2,  t}^*   =  0$; (ii) $p_{1,  st}^*  = 1$ and $p_{2,  st}^*  =  0$ or $p_{1,  st}^*  =  0$ and $p_{2,  st}^*  =  1$ and (iii) $p_{1,nt}^*    =(1 + \epsilon_1)/(1 - \epsilon_1 \epsilon_2)$ and $ p_{2,nt}^*   =  (1 + \epsilon_2)/(1 - \epsilon_1 \epsilon_2)$. 
These asymptotic solutions characterize the system according to their stability. 
The stability matrix, also called \textit{community matrix} \cite{murray},  is:
$A(p_1^*,p_2^*)  =  \left( \begin{array}{cc}
      \partial_{p_1} f & \partial_{p_2} f \\
      \partial_{p_1} g & \partial_{p_2} g   
      \end{array} \right)_{p_1^*,p_2^*}$.
The steady state solutions $p_1^*$ and $p_2^*$ are stable if the trace and the determinant of the community matrix are negative and positive, respectively. 
One has: $\mbox{Tr}[A(p_1^*,p_2^*)]  =  1 + \rho + ( \rho \epsilon_2 -2)p^*_1   
+ (\epsilon_1 - 2\rho) p^*_2$ and $\mbox{Det}[A(p_1^*,p_2^*)]  =  \rho \{  1+ p^*_1 [ \epsilon_2 -2(1+\epsilon_2 p^*_1)] +   p^*_2 [ \epsilon_1 -2 (1+ \epsilon_1 p^*_2 )] + 4 p^*_1 p^*_2 \}$. 
One must analyze the possible cases given $p^*_1$ and $p^*_2$. 

\subsection{Stability analysis}

Let us start with the stability analysis of the trivial solutions of $p_{1,  t}^*    = 0 $ and $ p_{2,  t}^*   =  0$, which means extinction of both species (synnecrosis). 
One has: $\mbox{Tr}   [A(0,0)]  =  1 + \rho$ and $\mbox{Det}[A(0,0)]  =  \rho$.
Since $\rho > 0$ , $\mbox{Det}[A(0,0)] > 0$ but $\mbox{Tr}[A(0,0)]  > 1$. 
The pair of trivial solution is not stable anywhere in the parameter space, so \textit{synnecrosis} never occurs in our model. 


The semi-trivial solutions are given by  $p_{1,  st}^*  = 1$ and $p_{2,  st}^*  =  0$ or $p_{1,  st}^*  =  0$ and $p_{2,  st}^*  =  1$ and they mean that one of the species is extinguished.  
Considering the species 1 extinction, one has: $\mbox{Tr}  [A(0,1)]  =  1+ \epsilon_1 - \rho$  and $\mbox{Det}[A(0,1)] =  - \rho (1+ \epsilon_1) $.
For these solutions to be stable,  it is necessary that $\epsilon_1<-1$, regardless of the $\rho$ value. 
A similar analysis leads us to conclude that species 2 extinction is stable only for $\epsilon_2<-1$.


The non-trivial solutions $p_{1,nt}^*    =(1 + \epsilon_1)/(1 - \epsilon_1 \epsilon_2)$ and $ p_{2,nt}^*   =  (1 + \epsilon_2)/(1 - \epsilon_1 \epsilon_2)$ lead to:
$\mbox{Tr}  [A(p_{1,nt}^*,p_{2,nt}^*)]  =  [1+ \epsilon_1 + (1 + \epsilon_2) \rho ]( \epsilon_1 \epsilon_2 -1)$ and $\mbox{Det}[A(p_{1,nt}^*,p_{2,nt}^* )]  =  - (1 + \epsilon_1)( 1+ \epsilon_2) \rho/(\epsilon_1 \epsilon_2-1)$. 
On one hand, if $\epsilon_1 \epsilon_2 < 1$, the  denominator is positive and the numerator of $p_{1,nt}$ and $p_{2,nt} $ must vanish or be positive.
From the condition $p_{1,nt}^* \ge 0 $, this solution is only stable if $\epsilon_1 \ge \epsilon^{(c)}_1 = -1$; otherwise, $p_{1,t}^*  =  0$ is the stable solution. 
This produces a transition from the regime where species 1 coexists with species 2 to the regime where species 1 is extinguished (Gause transition). 
The same transition occurs for the parameter $\epsilon_2$. 
On the other hand, if $\epsilon_1 \epsilon_2 > 1$, the denominator is positive and the numerator of $p_{1,nt}$ and $p_{2,nt} $ must be non-negative.
From the condition, $p_{1,nt}^* \ge 0 $, this solution is only stable if $\epsilon_1 <  \epsilon^{(c)}_1 = -1$; otherwise, $p_{1,t}^*  =  0$ is the stable solution. 
This produces a transition from the regime where species 1 coexists with species 2 to the regime where species 1 is extinguished. 
The same transition occurs for $\epsilon_2$. 

From the stability criteria, species can coexist only if $\epsilon_1 > - 1$ and $\epsilon_2 > - 1$.  
According to the values of $\epsilon_1$ and $\epsilon_2$, the various ecological regimes may present a stable non-trivial solution, as in \Fref{phase-diagram}. 

\subsection{Phase diagram}

\begin{figure}[htb]
 \begin{center}
 \includegraphics[width=\columnwidth]{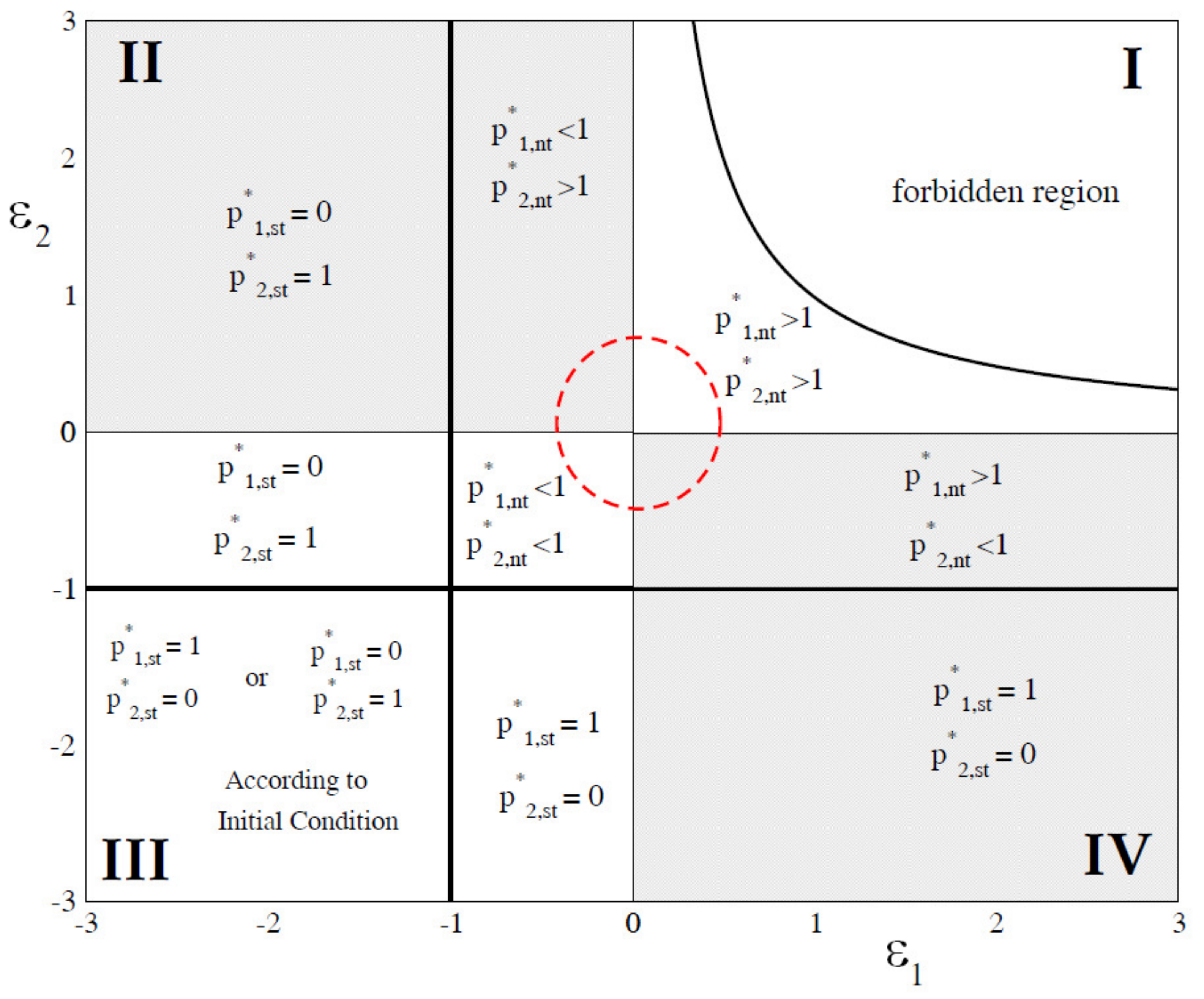}
 \end{center}
 \caption{Diagram of ecological interaction between two species according to the pair ($\epsilon_1,\epsilon_2$). 
               In this diagram, each quadrant represents one kind of interaction: I:mutualism; II and IV: predation ; III: competition. 
               The abscissa and ordinate represent either ammensalism or commensalism.
               The origin represents neutralism.
               The non-trivial solutions ($\epsilon_1>-1$, $\epsilon_2>-1$, and $\epsilon_1 \epsilon_2<1$) correspond to the coexistence phase.  
               In this phase mutalism, comensalism, amensalism, predation, competition, and neutralism can occur, for instance, around and inside the dashed circe. 
               The complementary region is characterized by an extinction phase and a forbidden region. 
               The extinction phase ($\epsilon_1 < -1$ and/or $\epsilon_2 < -1$) reveals a region for $\epsilon_1 < 0$ and $\epsilon_2 < 0$ where, contrary to the other cases, the steady state solutions depend on the initial condition. 
               For $\epsilon_1 > 1$ and $\epsilon_2 > 0$ and $\epsilon_2 > 1/\epsilon_1$, there exists a forbidden region with non biological reality (negative number of individuals). 
 }
 \label{phase-diagram}
 \end{figure}

In \Fref{phase-diagram}, the stable steady state solutions of \Eref{simple-model_1} and \Eref{simple-model_2} are represented in the parameter space. 
This diagram presents the reflexion symmetry about  $\epsilon_2 = \epsilon_1$ and summarizes our findings: the coexistence phase and one species extinction phase can be seen.
These phases span on different ecological regimes, which means that different ecological interactions may lead to the same phase. 

\begin{figure}[htbp]
 \begin{center}
 \includegraphics[width=\columnwidth]{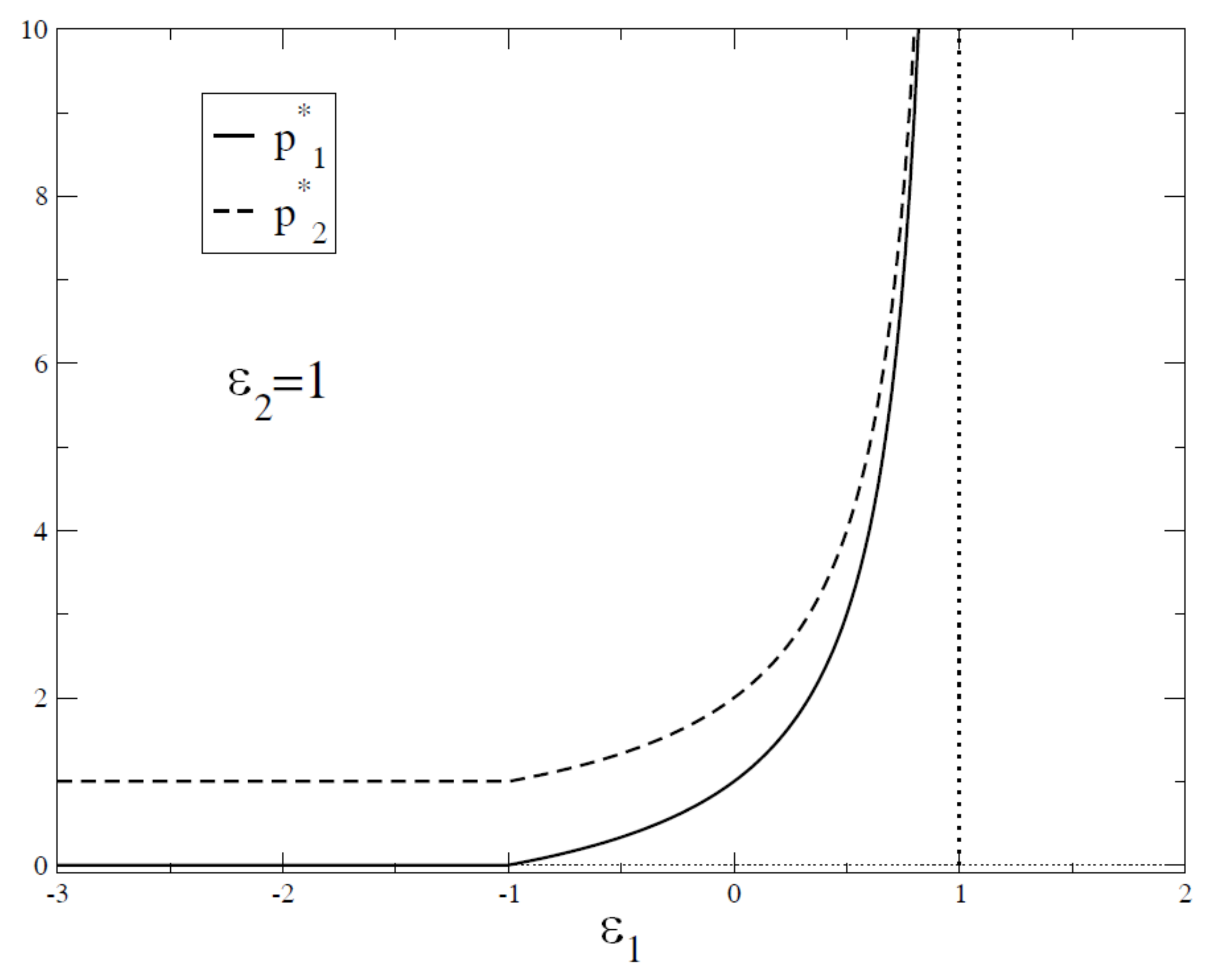}
 \end{center}
 \caption{Plot of steady steady solutions $p_{1,nt}^*    =(1 + \epsilon_1)/(1 - \epsilon_1 \epsilon_2)$ and $p_{2,nt}^*   =  (1 + \epsilon_2)/(1 - \epsilon_1 \epsilon_2)$ of the model [\Eref{simple-model_1} and \Eref{simple-model_2}] as a function of $\epsilon_1$, keeping  $\epsilon_2 = 1$ fixed.  
              In these plots one sees two phase transitions. 
              The first one is the extinction-coexistence (predation) transition, which occurs at $\epsilon_1 = -1$. Near $\epsilon_1^c$, the critical point,    the species 1 go to extinction by a linear form, that is  $p_{1,nt}^* \sim (\epsilon_1 - \epsilon_1^c)$.   The second one is the transition from coexistence (mutualism) to a forbidden region (with no ecological reality), which one with critical exponent $\beta = 1$. 
 }
 \label{fatia}   
 \end{figure}

For $\epsilon_1>-1$, $\epsilon_2>-1$, the non-trivial solutions are stable. 
They span on four considered ecological regions, around the origin of the phase diagram [see the circle in \Fref{phase-diagram}].
For mutualism, the first quadrant of the diagram of~\Fref{phase-diagram}, as $\epsilon_1 \epsilon_2 \to 1^-$, the mutual cooperation conducts to unbounded growth of both populations, so that $p_{i,nt} \sim  \left(  1- \epsilon_1 \epsilon_2 \right)^{-\beta}$ diverges with the exponent $\beta= 1$ (see~\Fref{fatia}). 
The region $\epsilon_2 > 1/\epsilon_1$ is forbidden;  since $p_{i,nt}^* < 0 $, it does not have ecological reality. 

The \Fref{fatia} also tell us that there is a transition between extintion and coexistence regimes. Keeping $\epsilon_2$ fixed, this transition occurs at $\epsilon_1 = \epsilon_1^c =-1$. A Taylor expansion of the non-trivial solution allow us to write $p_{1,nt}^* = (\epsilon_1 - \epsilon_1^c)/(1+ \epsilon_2) + {\cal O} \Big((\epsilon_1 - \epsilon_1^c)^2 \Big)$ , whith means that near $\epsilon_1^c$ (the critical point) species 1 linearly goes extinct, that is  $p_{1,nt}^* \sim (\epsilon_1 - \epsilon_1^c)$. The critical exponent related to the order parameter is $\beta=1$
Analogous process happens to the species 2.

For $\epsilon_1\le -1$ or $\epsilon_2 \le-1$ and $\epsilon_1 \epsilon_2<1$, the remaining regions of the diagram are characterized by the stability of the semi-trivial solutions. 
For $\epsilon_1<-1$ and $\epsilon_2<-1$, differently from all the other regions in the phase diagram, the steady state solutions depend on the initial condition. 
There is a separatrix for the initial conditions, in this region~\cite{murray}.

\subsection{Cancer modeling}
In cancer therapy, tumor-selective replicating viruses offer remarkable advantages over conventional therapies and are a promising new approach for human cancer treatment.
An oncolytic virus is a virus that preferentially infects and lyses cancer cells. 
Theoretical models of the interaction between an oncolytic virus and tumor cells are attainable using adaptations of techniques employed previously for modeling other types of virus-cell interaction~\cite{nowak}. 
A Lotka-Volterra like model that describes interaction between two types of tumor cells (the cells that are infected by the virus and the cells that are not infected but are susceptible to the virus so far as they have the cancer phenotype) and the immune system has been presented by Wodarz \cite{wodarz_2001,wodarz_2005}. 
This model can be written as \cite{novozhilov_2006}: $dx/d\tau=x\left[1-(x+y)\right]-\beta x y$; $dy/d\tau=\gamma y\left[1-(x+y)\right]-\beta x y-\delta y$,where, $x$ and $y$ are the population of uninfected and infected cells respectively, $\gamma$ is the ratio between infected and uninfected cells growth rates, $\beta$ is related to the interaction parameter between uninfected and infected cells and $\delta$ is related to the rate of infected cell killing by the virus.
Using \Eref{simple-model_1}, \Eref{simple-model_2} and making: $\epsilon_1=-(1+\beta)$, $\epsilon_2=(\beta/\gamma-1)$  and $\gamma \gg \delta$, one retrieves Wodarz model. 
Besides being simpler, our model presents all the qualitative behavior shown by Wodarz model, i.e. four different scenarios for the asymptotic states: (i) absence of infected cells, (ii) absence of uninfected cells, (iii) coexistence of both types of cells and (iv) dependence on initial conditions (all cells infected or uninfected).

\section{Model analytical solution}
\label{neutralism-amensalism-comensalism}

Here we present the analytical solution of the Verhulst-like Lotka-Volterra model. 
We start presenting the known solutions when one of the interacting parameter vanishes. 
Next we present the solution for the complete model.

\subsection{Vanishing of one interaction parameter ($\epsilon_1 \epsilon_2 = 0$)}

This section is restricted to the particular case $\epsilon_1 \epsilon_2 = 0$, where one or both interaction parameters vanish. 
This corresponds to the axis of the parameter space $\epsilon_2 \times \epsilon_1$ [see \Fref{phase-diagram}]. 
Thus, three ecological regimes are allowed in this specific situation: \textit{amensalism}:  $\epsilon_1 = 0$ and $\epsilon_2 < 0$ (species 2 extinction, if $\epsilon_2 \le -1$ and species coexistence otherwise) or $\epsilon_2 = 0$ and $\epsilon_1 < 0$ (species 1 extinction, if $\epsilon_1 \le -1$ and species coexistence otherwise); \textit{neutralism}:  $\epsilon_1 = \epsilon_2 = 0$;  and \textit{comensalism}:  $\epsilon_1 > 0$ and $\epsilon_2 = 0$ or $\epsilon_1 = 0$ and $\epsilon_2 > 0$. 

In these cases, one can obtain a simple full analytical solutions of \Eref{simple-model_1} and \Eref{simple-model_2}. 
Below we address each case in more detail.

\subsubsection{Neutralism}

Consider a special case where each population grows independent from the other. 
This ecological regime is represented by \Eref{simple-model_1} and \Eref{simple-model_2}, with $\epsilon_1 = \epsilon_2 = 0$,  leading to two independent Verhulst models: $ d p_1 / d\tau =   p_1[ 1 -p_1 ]$ and $ d p_2/(\rho d\tau) =   p_2[ 1- p_2 ]$.
The solutions, with different time scales and parameters, for each species are~\cite{brenno_2011}:  $p_{1}(\tau)  = 1/[1+(p_{1,0}^{-1}-1)e^{-\tau}]$  and $p_{2}(\tau)  = 1/[1+(p_{2,0}^{-1}-1)e^{-\rho \tau}]$, where $p_{i,0} = p_i(0)$ is the initial condition for species $i = 1,2$. 

The Verhulst solutions are driven by different characteristic times $\tau = \kappa_1 t$ and $\tau'= \rho \tau = \kappa_2 t$, respectively. 
For $\tau \gg 1$, so that $\rho \tau \gg 1$, the asymptotic behaviors $p_1^* = p_1(\infty) = 1$ and $p_2^* = p_2(\infty) = 1$ are obtained, so that species end exploring all the available environmental resources. 
The case $\epsilon_2 = 0$ in \Fref{com-ame-p2-x-tau} shows the dynamics of the population given by the Verhulst solutions. 
For $\kappa_2 > \kappa_1$; i.e., $\rho >1$,  species 2 grows more rapidly than species 1, given the same initial condition. 
For $\kappa_2 < \kappa_1$; i.e., $\rho <1$, the inverse occurs. 
 
\subsubsection{Comensalism and amensalism}

Consider that two species interact asymmetrically. 
For instance, consider that individuals of species 1 are unaffected by species 2, although, individuals  of species 2 are adversely affected by species 1. 
This is the amensalism regime. 
The commensalism regime has the same structure as amensalism, except that one species is favorably affected by the other. 
These interactions can be mathematically represented by the following equations: 
\begin{eqnarray}
 \frac{d p_1(\tau)}{        d\tau} &=&  p_1(\tau) \left[ 1 - p_1(\tau)                                       \right] \label{verhulst-model_1} \\
 \frac{d  p_2(\tau)}{\rho d\tau} &=&  p_2(\tau) \left[ 1 - p_2(\tau)  + \epsilon_2 p_1(\tau) \right] \label{verhulst-scha} \; ,
\end{eqnarray}
where $\epsilon_2$ is negative for amensalism and positive for commensalism. 

\begin{figure}[htbp]
 \begin{center}
 \includegraphics[width=\columnwidth]{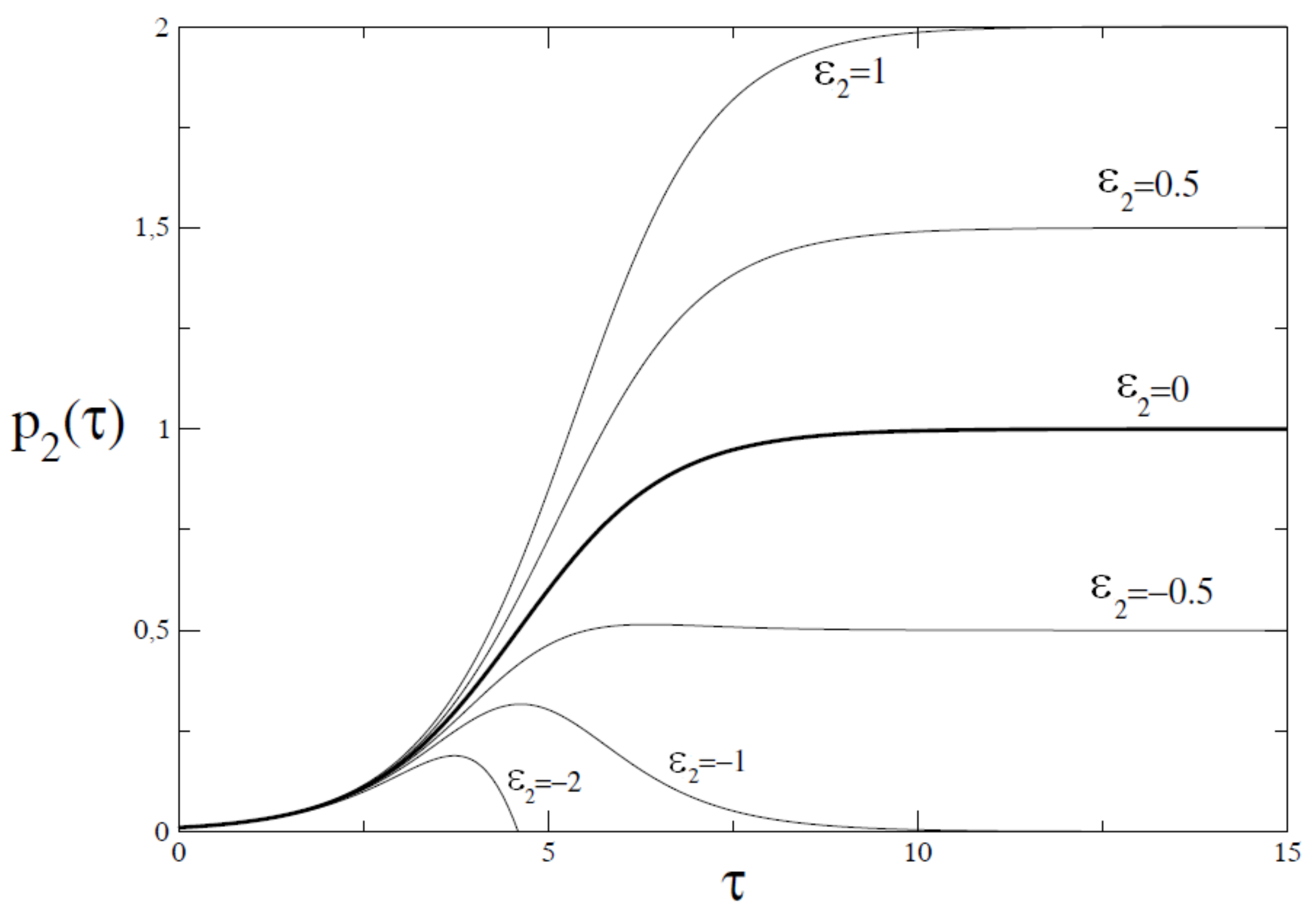}
 \end{center}
 \caption{Plots of the solution of \Eref{verhulst-scha}, given by  \Eref{eq:sol_time_verhulst_shaefer_model_3}, for different values of interaction parameter $\epsilon_2$ and $\rho = 1$.   
 The comensalism regime is obtained for $\epsilon_2 > 0$, where the asymptotic value, due to the other species, is greater than unity. 
 The neutralism regime is retrived for $\epsilon_2 = 0$. 
 The amensalism regime is obtained for $\epsilon_2 < 0$, where the asymptotic value does not vanish (meaning species coexistence) for 
 $\epsilon_2 > -1$ and vanishes (meaning species extinction) for $\epsilon_2 \le -1$.}
  \label{com-ame-p2-x-tau}  
 \end{figure}

In this kind of interaction, species 1, described by \Eref{verhulst-model_1}, follows the Verhulst model, whose solution is $p_{1}(\tau)  = 1/[1+(p_{1,0}^{-1}-1)e^{-\tau}]$ . 
The dynamics of species 2 follows the time-dependent Verhulst-Schaefer model [\Eref{verhulst-scha}], whose solution is~\cite{brenno_2011}: 
\begin{eqnarray}
\frac{1}{p_2(\tau)} & = &\frac{1}{1+\epsilon_2 p_1(\tau)}+ 
\frac{e^{-\rho\tau\left[1+\epsilon_2 \overline{p}_1(\tau)\right]}\left(1-p_{2,0}+\epsilon_2 p_{1,0}\right)}{p_{2,0}+\epsilon_2 p_{1,0} p_{2,0}}
\label{eq:sol_time_verhulst_shaefer_model_2} 
\end{eqnarray}
where  the mean relative size of species 1 up to $\tau$ is: 
\begin{eqnarray}
\fl 
\overline{p}_1(\tau) & = & \frac{1}{\tau} \int_0^{\tau} d \tau' p_1(\tau')  =  \int_{0}^{\tau}  \frac{d \tau'}{1 +  ( p_{1,0}^{-1} - 1) e^{-\tau'}} 
                                        =     \ln[1 + p_{1,0} (e^{\tau} - 1) ] \; .
\label{eq_integral_verhulst_amensalism}
\end{eqnarray}
Using the Verhulst solution for $p_1(\tau)$ and put the \Eref{eq_integral_verhulst_amensalism} in~\Eref{eq:sol_time_verhulst_shaefer_model_2}, one obtains:
\begin{eqnarray}
\nonumber
\frac{1}{p_2(\tau)} & = & \frac{e^{-\tau}\left[1+\left(-1+e^{\tau}\right)p_{1,0}\right]^2}{p_{1,0}\left\{1+\left[-1+e^\tau\left(1+\epsilon_2\right)\right]p_{1,0}\right\}}+ \\
&& \frac{e^{-\rho\tau}\left[1+\left(-1+e^{\tau}\right)p_{1,0}\right]^{-\epsilon_2\rho\tau}\left(1+\epsilon_2p_{1,0}-p_{2,0}\right)}{p_{2,0}+\epsilon_2p_{1,0}p_{2,0}}
\label{eq:sol_time_verhulst_shaefer_model_3} 
\end{eqnarray}
The plots of $p_2(\tau)$ for several $\epsilon_2$ values are depicted in \Fref{com-ame-p2-x-tau}.
 
The steady state solutions of \Eref{verhulst-model_1} and \Eref{verhulst-scha} are, respectively: $p_1^* = p_1(\infty)  = 1$ and $p_2^* = p_2(\infty)  = 1 + \epsilon_2$, if  $\epsilon_2 > -1$ or vanish otherwise. 
One sees that $\epsilon_2^{(c)} = -1$ is a critical value that separates two distinct phases: $\epsilon_2 \le -1$, where species 2 is extinguished; and  $\epsilon_2 > -1$, where species 2 coexists with species 1. 
The former case occurs in the amensalism regime, while in the latter one it may occur  in the amensalism  ($\epsilon_2 < 0$), neutralism ($\epsilon_2 = 0$), or in the comensalism ($\epsilon_2 > 0$) regimes. 

The same conclusions are valid for $\epsilon_2 = 0$ and $\epsilon_1 \ne 0$. 
One finds the same behaviors and a critical point $\epsilon_1^{(c)} = -1$, so that  similarly  for $\epsilon_1 < \epsilon_1^{(c)}$ species 1 is extinguished. 

\subsection {Mutualism, competition and predation ($\epsilon_1 \epsilon_2 \ne 0$)}
\label{numerical_solutions}

In the following, we deal with the case $\epsilon_1 \epsilon_2 \ne 0$, which addresses mutualism, competition, and predation. 
If 
\begin{itemize}
\item $\epsilon_1 \epsilon_2 >0$, each species has the same kind of influence on the other. 
This corresponds to either the competition or the mutualism regime. 
The following regimes occur: 
\begin{itemize}
\item $\epsilon_1 > 0$ and $\epsilon_2 > 0$, mutualism, which corresponds to the first quadrant of the space parameter phase space, restrict to the region $\epsilon_2 < 1/\epsilon_1$ (see \Fref{phase-diagram}; 
\item $\epsilon_1 < 0$ and $\epsilon_2 < 0$, competition, which corresponds to the third quadrant of the space parameter phase space (see \Fref{phase-diagram}).
 \end{itemize}
\item If $\epsilon_1 \epsilon_2 < 0$, the predation regime occurs, which belongs to the second and fourth quadrants of the parameter space (see \Fref{phase-diagram}).
\end{itemize}
For $\epsilon_2 > 0$ and $\epsilon_1 < 0$, there is species coexistence for $\epsilon_1 > \epsilon_1^{(c)} = -1$ and species 1 extinction for $\epsilon_1 \le \epsilon_1^{(c)}$. 

These ecological regimes are special cases  of \Eref{simple-model_1} and \Eref{simple-model_2}, whose solutions can be worked out to have the form:
\begin{eqnarray}
\frac{1}{p_1(\tau)} & = & \frac{1}{1+\epsilon_1 p_2(\tau)} + 
\frac{e^{-\tau\left[1+\epsilon_1 \overline{p}_2(\tau)\right]}\left(1-p_{1,0}+\epsilon_1 p_{2,0}\right)}{p_{1,0}+\epsilon_1 p_{1,0} p_{2,0}}
\label{eq:sol_time_verhulst_shaefer_model_1b}
\end{eqnarray}
where $p_2(\tau)$ is given by \Eref{eq:sol_time_verhulst_shaefer_model_2}, and the relative populations sizes mean values up to instant $\tau$ are $\overline{p}_1(\tau)  =   \int_0^{\tau} d \tau' p_1(\tau')/\tau$ and $\overline{p}_2(\tau)  =   \int_0^{\tau} d \tau' p_2(\tau')/\tau$.

\begin{figure}[htbp]
 \begin{center}
 \includegraphics[width=\columnwidth]{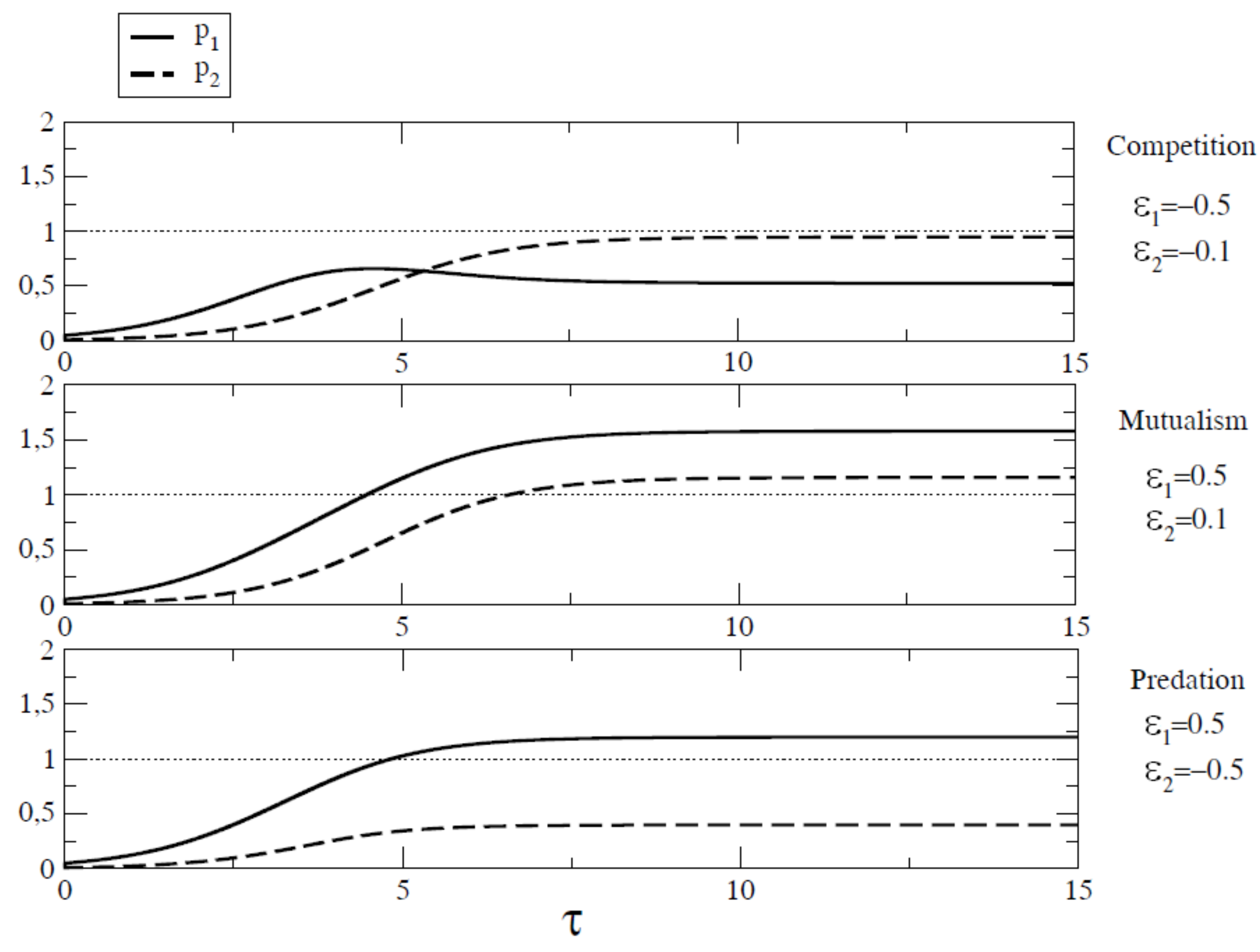}
 \end{center}
 \caption{Plots of the evolution equations \Eref{eq:sol_time_verhulst_shaefer_model_1b} and \Eref{eq:sol_time_verhulst_shaefer_model_2}, which are solutions of \Eref{simple-model_1} and \Eref{simple-model_2}. The initial condition is $p_{1,0} = p_{2,0} = 1/100$ and $\rho = 1$. 
 The species interaction parameters $\epsilon_1 \epsilon_2 \ne 0$ so that for competition: $\epsilon_1 = -1/2$ and $\epsilon_2 = -1/10$; mutualism: $\epsilon_1 = -1/2$ and $\epsilon_2 = -1/10$; and predation: $\epsilon_1 = 1/2$ and $\epsilon_2 = -1/2$.}
 \label{numerical-graph}
 \end{figure}

Using \Eref{eq:sol_time_verhulst_shaefer_model_2} in \Eref{eq:sol_time_verhulst_shaefer_model_1b},  we obtain a quadratic equation for $p_1(\tau)$, eliminating its dependence on $p_2(\tau)$. In fact,  we can write $p_1(\tau)$ as dependent only on the initial condition and $\overline{p}_2(\tau)$. 
The population size $p_2(\tau)$ behaves analogously. 
Thus, the coupling between the two population sizes is given only by the mean values. 
The solutions \Eref{eq:sol_time_verhulst_shaefer_model_1b} and \Eref{eq:sol_time_verhulst_shaefer_model_2} are presented in \Fref{numerical-graph} for the three regimes where $\epsilon_1 \epsilon_2 \ne 0$. As $\tau \to \infty$, the steady state solutions of \Eref{simple-model_1}   and \Eref{simple-model_2} are reached.

With the analytical solution, one can access the transient behavior of a given population. Transient dynamics can be an important aspect of the coexistence of predators and preys, and also of competitors \cite{hastings_2004}. Studies of outbreaks (insects or diseases) focuses greatly on the transient dynamics \cite{gavrilets_1995,lai_1995,lai_1995_2,lai_1995_3,cavalieri_1995,harrison_2001,kaitala_1999}. In tuberculosis treatment for example, it can reveal important aspects beyond asymptotic states such as how drug resistance emerges \cite{espindola_2011}.
In the present study \Fref{numerical-graph} illustrates the importance of the transient in the time evolution of the species densities. 
Consider the case of competition, for $\tau<5$ species 1 population is greater than species 2, however the steady state solution for this system is just the opposite (species 2 population is greater than species 1).
In this case, a simple steady state analysis would not be coherent with reality if the observation time scale is not appropriate. In other words, if the observed system has not yet reach equilibrium, the steady state analysis can be misleading.

Notice that considering $\epsilon_1 = 0$ in \Eref{eq:sol_time_verhulst_shaefer_model_1b} and \Eref{eq:sol_time_verhulst_shaefer_model_2}, one retrieves the Verhulst solution and \Eref{eq:sol_time_verhulst_shaefer_model_2}, and \Eref{eq_integral_verhulst_amensalism}, which correspond to the amensalism, neutralism, and comensalism regimes. 
In this way, these evolution equations can be seen as a general solution that is valid for all kinds of interaction regime. 
 
\section{Conclusion}
\label{conclusion}

The simple model we addressed here illustrates that one can interpret the interaction of a two species system at several levels. 
From the interaction parameters $\epsilon_1$ and $\epsilon_2$, which act at the individual level of species, one is able to tell about the different  ecological regimes, classified in a higher level according to the product of the interaction parameter $\epsilon_1 \epsilon_2$. 
If it vanishes, one or two species are independent from each other. If  $\epsilon_1 \epsilon_2 >0$, one has either mutualism (both positive) or competition (both negative). For $\epsilon_1 \epsilon_2 < 0$, one has predation. 
A collective level is obtained from the stability of the steady state solution, from where one obtains three phases: extinction of one species ($\epsilon_i < -1$) (synnecrosis is not a stable phase in our model), species coexistence, and a forbidden phase ($\epsilon_2 > 1/\epsilon_1$). 
Although the studied model has been considered in several isolated instances, our study reveals the very general aspect of a simple mathematical set of equations, which represents very rich ecological scenarios that can be described analytically.
In this manuscript we focused on a Verhulst term for the population growth in the Lotka-Volterra equation. 
All the results presented here can be extended using a more general growth model, for instance the Richards' model~\cite{richards_1959,martinez:2008b,Martinez:2009p1410} and implications of such generalizations will be detailed in a brief future.

\ack

The authors thank Roberto Kraenkel for fruitful discussion and numerous suggestions on this issue. 
F. R. acknowledges support from CNPq (151057/2009-5). 
B. C. T. C. acknowledges support from CAPES.
A.S.M. acknowledges support from CNPq (305738/2010-0 and 476722/2010-1). 

\section*{References}


\end{document}